# Early Detection of Alzheimer's Disease using Bottleneck Transformers

Ananya Sadana[1] and Arunima Jaiswal[2]

*Department of Computer Science and Engineering, Indira Gandhi Delhi Technical University for Women, Delhi, India*
*ananya097btcse19@igdtuw.ac.in[1], arunimajaiswal@igdtuw.ac.in[2]*

*Abstract:* Early detection of Alzheimer's Disease (AD) and its prodromal state, Mild Cognitive Impairment (MCI), is crucial for providing suitable treatment and preventing the disease from progressing. It can also aid researchers and clinicians to identify early biomarkers and minister new treatments that have been a subject of extensive research. The application of deep learning techniques on structural Magnetic Resonance Imaging (MRI) has shown promising results in diagnosing the disease. In this research, we intend to introduce a novel approach of using an ensemble of the self-attention-based Bottleneck Transformers with a sharpness aware minimizer for early detection of Alzheimer's Disease. The proposed approach has been tested on the widely accepted ADNI dataset and evaluated using accuracy, precision, recall, F1 score, and ROC-AUC score as the performance metrics.

*Keywords* – Alzheimer's Disease, Mild Cognitive Impairment, Magnetic Resonance Imaging, Computer-Aided Diagnosis, Deep Learning, Self-Attention, Bottleneck Transformers

## 1. Introduction

Alzheimer's disease (AD) is a widely prevalent neurodegenerative disease of the elderly population in the world. There are more than 55 million dementia patients in the world according to the World Health Organization. 139 million people are predicted to be living with dementia by 2050, adding a substantial burden to the economy, healthcare system, and society in general. In its current form, AD constitutes about 60-70% of all dementia cases (World Health Organization, 2021). There is currently no available treatment that can cure AD or completely prevent its progression. There is ample research being carried out to identify the cause of the disease, yet, the exact cause has not been able to be inferred. At an early stage, it may appear as normal episodes of forgetfulness that most overlook thinking it to be attributed to old age. However, it gradually worsens to a state where the person is unable to perform even basic cognitive tasks and requires constant supervision. Mild cognitive impairment (MCI) is the bridge between the clinically diagnosed AD and the expected normal aging with regards to cognitive abilities. MCI patients are more likely to develop AD as compared to the healthy cognition belonging to the same age group (Liu et al., 2014). Hence, it is crucial to detect AD and MCI at their earliest possible stage and prevent them from progressing.

There are various diagnostic measures used for identifying AD. Some of them include measurement of amyloid, tau, and cerebrospinal fluid (CSF), neuroimaging studies, like Magnetic Resonance Imaging (MRI), and Positron Emission Tomography (PET), and neurogenetic diagnostics (Salvatore et al., 2015). MRI is a neuroimaging modality that uses radio waves and strong magnetic fields to provide a three-dimensional view of the brain without invasiveness. Through this technology, the morphology of the human brain *in vivo* has been able to be investigated (Sabuncu et al., 2015). Therefore, allowing visualization of underlying AD-related changes in the brain. In practice, these scans are interpreted by radiologists and physicians mostly for diagnosing AD. However, when the disease is in its early stages it is at times harder for humans to be able to detect accurately from images. To tackle this problem, computer-aided diagnosis has begun to be used by researchers and doctors to assist them in diagnosing more accurately.

As a tool for computer-aided diagnosis of a variety of diseases, deep learning in the field of medical image analysis has proven extremely effective. To detect AD from brain MRI scans, several deep learning techniques have been employed, including deep belief networks (DBNs) (Brosch et al., 2013), stacked autoencoders (SAEs) (Gupta et al., 2013), and convolutional neural networks (CNNs) (Hosseini-Asl et al., 2016). With respect to vision tasks like image segmentation, classification, and object detection, CNNs are the most widely used architecture. This is due to the fact that they can directly accept image data as input and can effectively capture local information from them. CNN models trained on MRI scans can also automatically retrieve features for the learning process and thus, obviating the use of manual feature selection (Li and Liu, 2018). However, convolutions alone are not effectively able to model long-range dependencies, and to globally cumulate local filter responses, stacking multiple layers is needed. This is something that self-attention models like transformers are capable of dealing with. Transformers (Vaswani et al., 2017) have revolutionized the field of natural language processing (NLP) in the last few years due to their ability to capture long-range dependencies which is a critical characteristic in NLP. These ideas have steadily started being implemented in the computer vision domain as well. Pure Attention models (like Stand-Alone Self-Attention (SASA) model (Ramachandran et al., 2019), Local Relation Networks (LRNet) (Hu et al., 2019), Self-Attention Network for image recognition (SANet) (Zhao et al., 2020), Vision Transformer (ViT) (Dosovitskiy et al., 2021), etc.) and hybrid Convolution + Attention models (like Attention Augmented Convolutional Networks (AA-ResNet) (Bello et al., 2019), Video – Bidirectional Encoder Representations from Transformers (VideoBERT) (Sun et al., 2019), Criss-Cross Attention for semantic segmentation (CCNet) (Huang et al., 2019), Detection Transformer (DETR) (Carion et al., 2020), etc.) have started gaining popularity and are being applied to various image classification and segmentation tasks. One such hybrid model is the Bottleneck Transformer (BoTNet) (Srinivas et al., 2021) that has been used in this research.

The analysis of structural MRI scans of the brain in this paper was designed to identify subjects with MCI and AD at an early stage, by distinguishing between three categories (1) AD and CN (cognitively normal), (2) MCIc (patients with MCI converting to AD within a span of 18 months) and CN, and (3) MCIc and MCInc (patients with MCI not converting to AD within a span of 18 months). Patients who fall under the incubation and the illness period of AD are often diagnosed with MCI; however, not all of the subjects with MCI diagnosis make a transition to AD (Salvatore et al., 2015). As a result, it is imperative to distinguish between MCI converts and MCI non-converters so that the transition from mild or asymptomatic dementia to AD can be detected early and treated accordingly. The classification was made using BoTNet (Srinivas et al., 2021) as the base classifier in an ensemble of central 2D slices obtained from the brain scans. The performance of this model was then evaluated on the basis of accuracy, precision, recall, F1 score, and ROC-AUC score.

The paper has been assembled in the following manner. Section 2 provides a brief about the background of the problem statement being dealt with by analyzing the related studies available in the literature. This is then followed by section 3 that describes the entire architecture of the research work in detail. It is divided into three subsections explaining the dataset used, data preprocessing, and finally the model architecture. The next section covers an analysis of the results obtained from the implemented work. Lastly, section 5 provides a conclusion and discusses the future scope of this work.

## 2. Background work

Neuroimaging has started gaining some substantial popularity in recent times for accurately observing physical changes in the brain causing AD. From various neuroimaging methods, such as functional magnetic resonance imaging (fMRI), structural magnetic resonance imaging (sMRI), single-photon emission computed tomography (SPECT), etc., it is readily apparent that degenerating brain cells bring about these changes (Islam et al., 2018). In addition to aiding in early detection, they also aid in identifying specific regions of the brain most affected by the disease. The need for reliable tools and methods for processing neuroimaging data from these modalities is evident, therefore. Analysis of neuroimaging data has greatly benefited from the use of machine learning techniques. Many researchers have adopted these techniques and are building increasingly enhanced classifiers to detect AD (Liu et al., 2014; Salvatore et al., 2015; Islam et al., 2018; Li and Liu, 2018; Pan et al., 2020). As successfully identified by researchers, the hippocampus, the entorhinal cortex, the gyrus rectus, the basal ganglia, along other regions of the brain have been shown to undergo major structural changes during AD (Salvatore et al., 2015).

One of the most common approaches found in several research works was developing intricate machine learning models to help detect AD using MRI scans. Klöppel et al. (2008) used linear support vector machines (SVM) to detect AD by classifying the grey matter portion of brain MRI scans. Using MRI and fluorodeoxyglucose (FDG) - PET scans, Gray (2012) developed an AD classification model using a multi-modality approach based on random forest. Salvatore et al. (2015) first extracted and selected features from the whole brain using Principal Component Analysis or PCA pairing it with Fisher Discriminant Ratio or FDR criterion and then carried out binary classifications among the groups - AD, CN, MCIc, and MCInc by building a machine learning model based on SVM.

Deep learning frameworks have also risen in popularity as they are able to learn abstract feature representations from data. They are known to perform excellently in computer vision tasks and are increasingly being applied to medical imaging for the same. Brosch et al. (2013) used manifold learning based on deep belief networks (DBN) to detect AD from 3D MRI scans. Liu and Shen (2014) used unsupervised and supervised learning to learn deep features from MRI scans for AD and MCI classification based on a pre-trained CNN model. Pan et al. (2020) aimed at early detection of AD by combining CNNs with ensemble learning on MRI slices to perform the classifications as done by Salvatore et al. (2015). They also developed a 3D Squeeze-and-Excitation Networks model (3D-SENet) that makes use of a channel attention mechanism.

In this paper, a novel approach has been proposed to detect AD by making use of Bottleneck Transformers (Srinivas et al., 2021), a deep learning model, as the base classifier with a sharpness aware minimizer in an ensemble of T1 weighted MRI scans sliced along the coronal axis. The same classification approach has been followed as performed by Salvatore et al. (2015) and Pan et al. (2020) in their respective research works. This methodology of employing BoTNet for the early detection of AD is the first of its kind.

# 3. System architecture
## 3.1. Dataset

All the data used in this study has been obtained from the Alzheimer's Disease Neuroimaging Initiative (ADNI) database (http://adni.loni.usc.edu/). As a public-private partnership, ADNI was launched in 2004 under the leadership of Dr. Michael W. Weiner, with the goal of detecting Alzheimer's disease at its earliest stage. Using MRI, PET, and clinical and neuropsychological testing, ADNI examines the progression of early Alzheimer's disease and mild cognitive impairment.

This study examines the structural MRI brain scans with T1 weighting for a total of 569 subjects. Out of these 569 subjects, 201 subjects identified as cognitively normal elderly controls, 136 subjects were diagnosed with AD, 92 subjects with MCI diagnosis who converted to AD within a span of next 18 months (MCIc), and 140 subjects with MCI diagnosis but did not convert to AD within the span of next 18 months (MCInc). In this study, participants who had been followed for less than 18 months were excluded. Both baseline and non-baseline brain scans were included to facilitate better training. However, in an effort to prevent overlap between the training/validation and testing datasets, the datasets were constructed according to the IDs of the subjects.

*Table 1. Number of subjects and MRI scans for each category of classification*

| Category | Unique subjects | Total scans |
|---|---|---|
| CN | 201 | 764 |
| AD | 136 | 739 |
| MCIc | 92 | 376 |
| MCInc | 140 | 426 |

Cognitively normal subjects were defined as having Mini-Mental State Examination (MMSE) scores between 24 and 30 and Clinical Dementia Rating (CDR) (Morris, 1993) as zero along with the non-existence of any signs of depression, MCI, or dementia. In order to be considered for MCI, individuals had to have an MMSE score between 24 and 30 and a CDR of 0.5 with objective loss of memory, examined by measuring scores on the Wechsler Memory Scale Logical Memory II (Wechsler, 1987) along with the non-existence of major signs of dementia. Lastly, AD criteria for inclusion included MMSE scores between 20 and 26 and a CDR between 0.5 and 1; as well as the ADRDA/NINCDS standard for likely AD (Dubois et al., 2007; McKhann et al., 1984).

## 3.2. Data Processing

Post-processed T1 weighted MRI scans were downloaded from the ADNI database in *.nii* (NIfTI) format. The brain scans had a field strength of 1.5T and the plane of acquisition as Sagittal. The processing pipeline followed by ADNI first included the images to undergo B1 correction for non-uniformity (Narayana, 1988), and 3D GradWarp correction for non-linearity of gradient (Jovicich et al., 2006) using the UCSD package. Subsequently, the Scaled process and N3 ADNI pipeline were also implemented. Further processing of the MRI brain scans was carried out using

FreeSurfer's (http://surfer.nmr.mgh.harvard.edu/) cross-sectional and longitudinal processing pipeline. Finally, the cross-sectionally and longitudinally processed brain masks, registered to Talairach space, were collected for the purpose of this study.

In order to accelerate the training and testing of the model, the 3-dimensional brain scans were then sliced into 2-dimensional images along the coronal axis taking ten central brain slices from each scan. Figure 1 shows a few examples of the 2D slices of a subject's brain scan. The slices were then cropped to give images with a resolution of 224×224 and then normalized using min-max normalization. They were then flipped and translated along the sagittal axis to augment the data and avoid possible over-fitting.

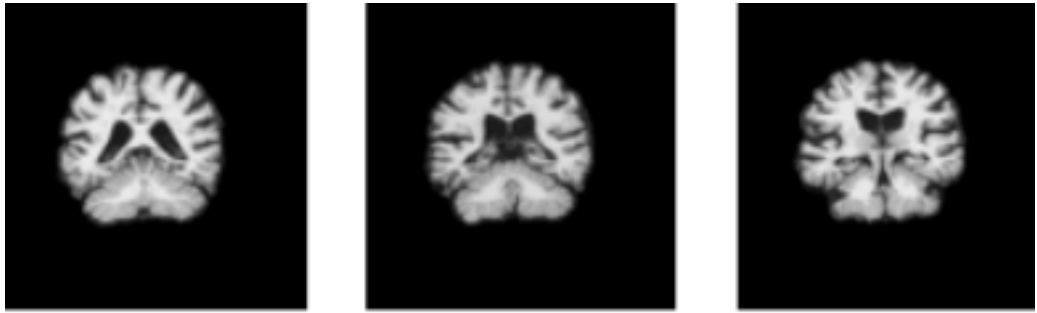

*Figure 1. Central brain slices from MRI scan*

### 3.3. Model Architecture

There were three binary classifications made namely: AD vs. CN, MCIc vs. CN, and MCIc vs. MCInc. All these classifications followed the same architecture of modeling i.e., training ten BoTNet-50 models corresponding to ten different central 2D brain slices. The training was carried out in a 5 folded cross-validation loop with 60 epochs per fold where the best model for each slice was chosen based on its validation accuracy. Figure 2 visually demonstrates the flow of work of the entire system architecture for each binary classification.

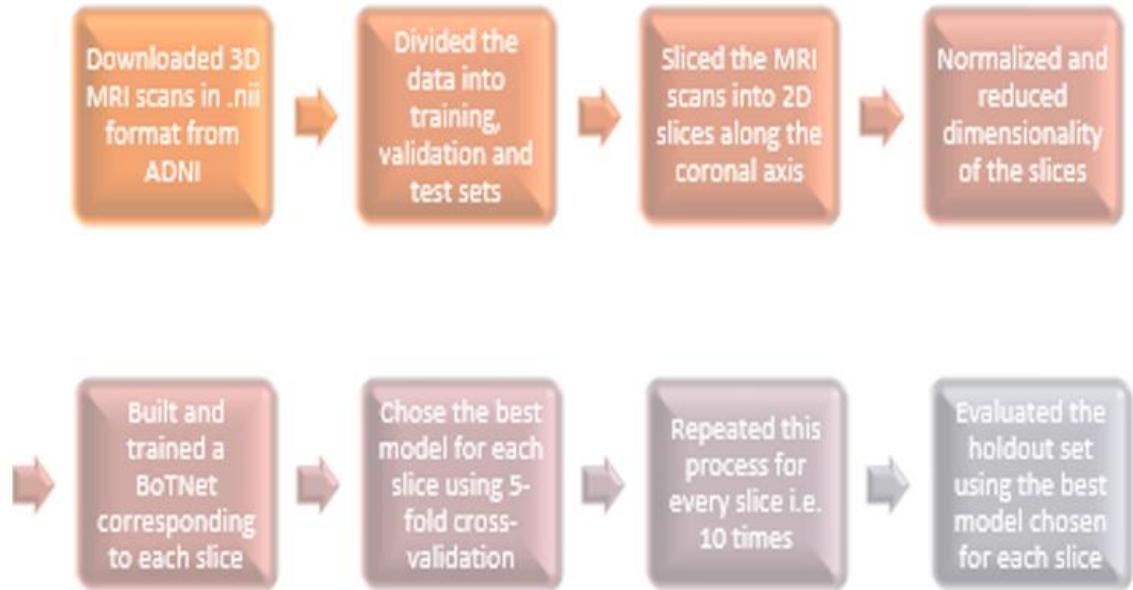

*Figure 2. The framework of each binary classification*

3.3.1. BoTNet

The architecture of BoTNet (Srinivas et al., 2021) has been correctly described as simple yet powerful by its creators. It differs from ResNet in just the way that the spatial 3×3 convolution layers in the last three bottleneck blocks are replaced with Multi-Headed Self-Attention (MHSA) layers. These blocks can be seen as Transformer (Vaswani et al., 2017) blocks.

Residual Net or ResNet (He et al., 2015) is a popular deep learning algorithm proposed in 2015 to tackle the problem of exploding/vanishing gradient associated with deep convolutional neural networks as the number of layers increases. A method known as identity shortcut connections was introduced that skips one or more layers and connects to the output directly. The idea is that instead of input 'x' being multiplied by weights of each layer and then adding a bias term, the network fits the residual mappings. Thus, the initial mapping H(x)=F(x) is transformed to H(x)=F(x)+x which is a relatively easier mapping to optimize.

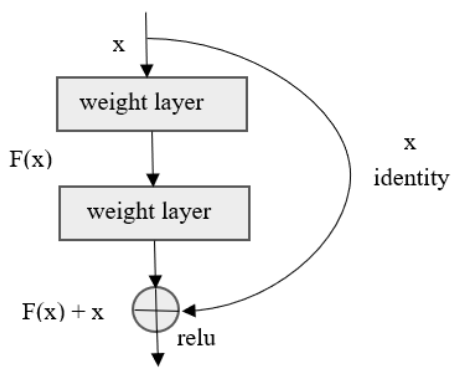

*Figure 3. Residual building block for ResNet*

BoTNet incorporates self-attention into ResNet's backbone architecture in a way that lets convolutions downsample first and then the self-attention mechanism takes over from there and works on the lower resolutions as seen in table 2 and figure 4. This approach significantly improves the performance of a typical ResNet. Just as in ResNet, BoTNet consists of four stages [C2, C3, C4, C5] having multiple bottleneck blocks. Following the architecture of ResNet-50, the model of BoTNet that is used in this research work also consists of [3, 4, 6, 3] stacks of bottleneck blocks in its last four stages respectively. Table 2 portrays the architecture of BoTNet-50 that is used here as described in the paper by Srinivas et al., 2021. Figure 5 illustrates the Bottleneck Transformer (BoTNet) block, which is the last block in the model.

*Table 2. The architecture of the BoTNet-50 model used*

| Stage | Size of the output | BotNet-50 |
|---|---|---|
| C1 | 112×112 | 7×7, 64, stride 2 |
| C2 | 56×56 | 3×3 max pool, stride 2 <br> $\begin{bmatrix} 1\times 1, & 64 \\ 3\times 3, & 64 \\ 1\times 1, & 256 \end{bmatrix} \times 3$ |
| C3 | 28×28 | $\begin{bmatrix} 1\times 1, & 128 \\ 3\times 3, & 128 \\ 1\times 1, & 512 \end{bmatrix} \times 4$ |
| C4 | 14×14 | $\begin{bmatrix} 1\times 1, & 256 \\ 3\times 3, & 256 \\ 1\times 1, & 1024 \end{bmatrix} \times 6$ |
| C5 | 7×7 | $\begin{bmatrix} 1\times 1, & 512 \\ MHSA, & 512 \\ 1\times 1, & 2048 \end{bmatrix} \times 3$ |

### 3.3.2. Multi-Head Self-Attention Layer

Attention is the technique that has the capability of selecting and focusing on only the most pertinent information. In self-attention essentially the inputs are able to interact with each other to identify what parts to pay more attention to. This mechanism employs similar pixels and disregards the ones having no correlation to the other pixels in a feature map (Yang, 2020). This works in a way that each attention head presides over an input sequence, $x = (x_1, x_2, \ldots, x_n)$ where $x_i \in \mathbb{R}^{d_x}$, and outputs another sequence $z = (z_1, z_2, \ldots, z_n)$ where $z_i \in \mathbb{R}^{d_z}$. Here $d_x$ and $d_z$ refer to the embedding size of input and output sequence respectively. The elements in the output sequence z are calculated as the weighted sum of linearly transformed input elements (Shaw et al., 2018).

$$z_i = \sum_{j=1}^{n} \alpha_{ij}(x_j W^V) - (1)$$

The weight coefficient α_{ij} is calculated with the help of a softmax function as follows:

$$\alpha_{ij} = \frac{\exp(e_{ij})}{\sum_{k=1}^{n} \exp(e_{ik})}$$

Here $e_{ij}$ is calculated by implementing a scaled dot product to compare two input values:

$$e_{ij} = \frac{(x_i W^Q)(x_i W^K)^T}{\sqrt{d_z}} - (2)$$

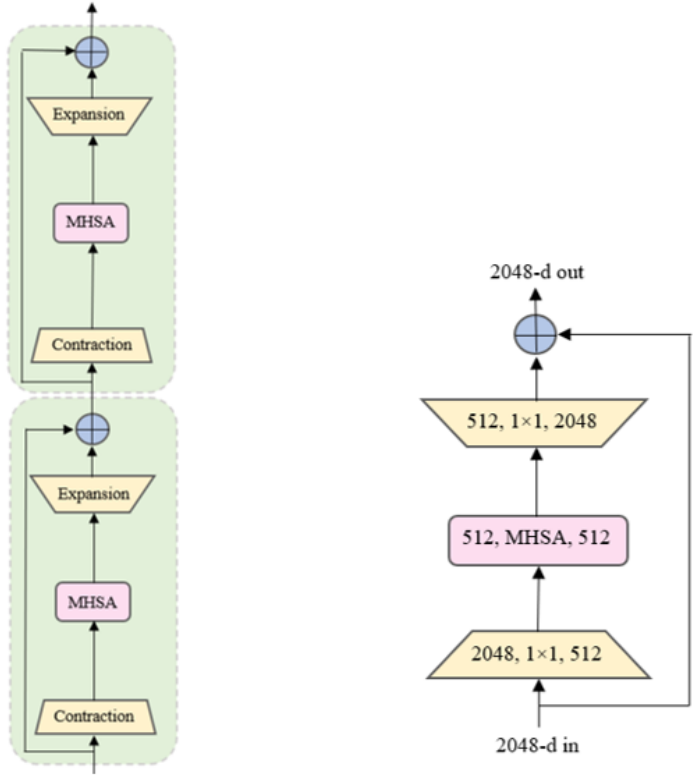

*Figure 4. Bottleneck Transformer (BoT) block of the model behaving as a ResNet bottleneck block in the architecture*

In multi-headed self-attention, individual attention heads are generally concatenated and run parallelly with the application of a linear transformation. This permits the model to together pay attention to information from multiple representation subspaces simultaneously (Vaswani et al., 2017). On a 2D feature map, global self-attention (all2all attention) in BoTNet is implemented with relative position encodings (Srinivas et al., 2021). Figure 5 provides a representation of this self-attention layer as used by Srinivas et al., 2021 in their model. In figure 5, $R_w$ and $R_h$ refer to the relative position encodings corresponding to the width and height respectively. $qr^T$ and $qk^T$ are the attention logits where q, k, and r represent query, key, and relative position encodings respectively. $W_Q$, $W_K$, $W_V \in \mathbb{R}^{d_x \times d_z}$ are projection matrices to create the query, key, and value vectors respectively. Every layer and attention head has its own projection matrix. 1×1 refers to a pointwise convolution, while element-wise summation and matrix multiplications are represented by ⊕ and ⊗ respectively. Transformer architectures often utilize position encoding to make attention mechanisms aware of position (Vaswani et al., 2017).

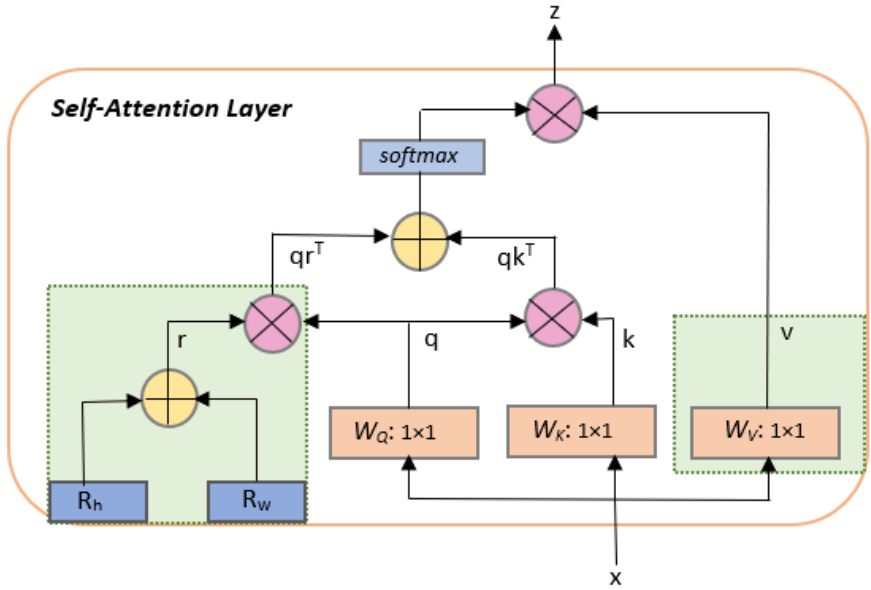

*Figure 5. Multi-Head Self-Attention (MHSA) layer used in the last block of BoTNet*

### 3.3.3. Relative Position Encodings

Recently it has been noticed that position encodings aware of relative distance perform better on computer vision tasks (Shaw et al., 2018). This can be credited to attention considering the relative distances between features at different locations and not just the content information. As seen from figure 5, BoTNet makes use of 2D relative position encoding that is able to viably connect information across objects with positional awareness (Srinivas et al., 2021). The use of relative position encoding can help save parameters as it can be shared among various heads (Wu et al., 2021). As proposed in (Shaw et al., 2018) and later improved by (Wu et al., 2021) the edge information or value embeddings in relation-aware self-attention can be calculated by modifying eq. (1) as follows:

$$z_i = \sum_{j=1}^{n} \alpha_{ij}(x_j W^V + r_{ij}^V) - (3)$$

where $r_{ij}^V \in \mathbb{R}^{d_z}$ is a vector depicting the edge between inputs $x_i$ and $x_j$ and $d_z = d_x$

Next, the equation to calculate the compatibility function i.e., eq. (2) is also modified as follows:

$$e_{ij} = \frac{(x_i W^Q)(x_i W^K)^T + b_{ij}}{\sqrt{d_z}} - (4)$$

where $b_{ij} \in \mathbb{R}$ is the 2D relative position encoding and is calculated as,

$$b_{ij} = (x_i W^Q) r_{ij}^T - (5)$$

where $r_{ij} \in \mathbb{R}^{dz}$ is a vector interacting with the query embedding

### 3.4. Hyperparameters

In this study, 8 heads for the MHSA layer of the BoTNet-50 model were used. Other hyperparameters that were included along with their optimized values are mentioned in Table 2. The loss function used is the Cross-Entropy Loss which is the most popular choice in deep learning. It computes the performance of the classification task by taking the sums of negative logarithms of probabilities of different classes. To avoid rigorous pretraining which is a common practice in models using the self-attention mechanism, the recently proposed sharpness aware minimizer (Foret et al., 2021) with a base optimizer as Adam was used. This method improved the generalization ability of the model by smoothing the loss function. The motivation behind the sharpness aware minimization is that instead of searching for parameter values that reduce the training loss, it looks out for values of parameters wherein their entire neighborhood has a consistently low training loss value. Thus, it simultaneously reduces the loss value and loss sharpness (Foret et al., 2021).

*Table 3. Hyperparameters of the model and their optimized values*

| Hyperparameter | Optimized value |
|---|---|
| Number of epochs | 60 |
| Learning rate | 3e-5 |
| Weight decay | 3e-5 |
| Loss criterion | Cross-Entropy Loss |
| Optimizer | SAM (with base optimizer as Adam) |

## 4. Result and analysis

The dataset was initially divided such that 80% of data went to the training set, 10% to the validation set, and an additional 10% was held out for the testing purpose to verify the efficacy of the proposed technique. The average classification accuracies after five-fold cross-validation on the ensemble of 10 different slices were 91.67% for AD vs. CN, 85.22% for MCIc vs. CN, and 82.67% for MCIc vs. MCInc. As seen from Table 4 these accuracies are significantly higher than the ones achieved by Salvatore et al., 2015 and Pan et al., 2020 using PCA+SVM and CNN-EL approaches respectively. Figure 6 displays the training and validation loss and accuracy as functions of epochs during training.

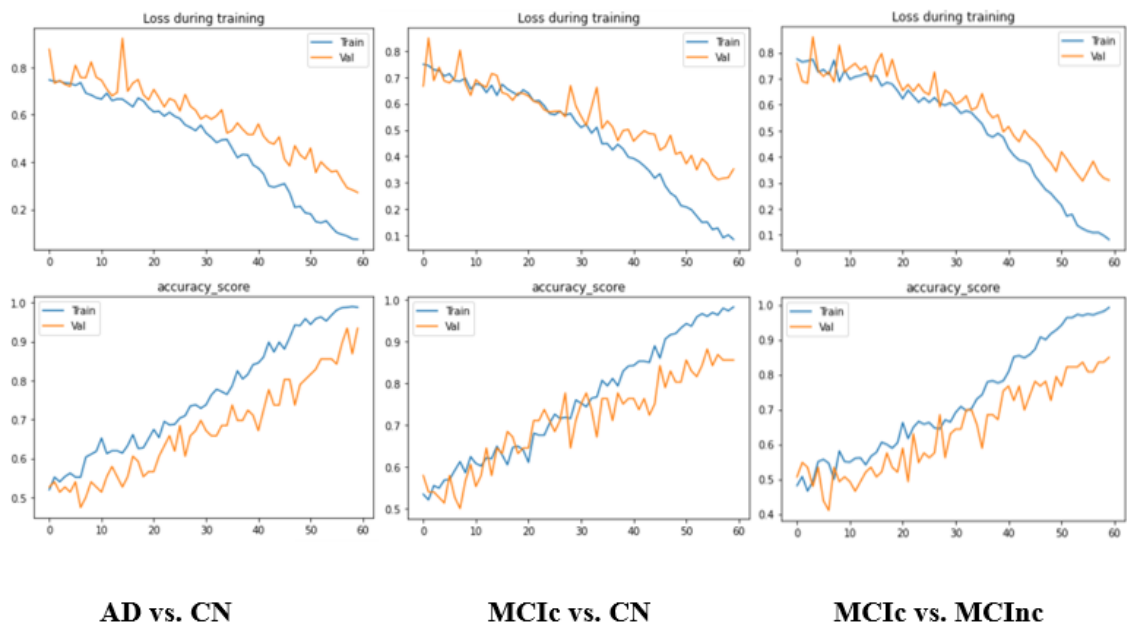

*Figure 6. Training versus Validation loss and accuracy plots for the three classifications*

To more extensively evaluate the performance of the model, Accuracy, Precision (or positive predictive value), Recall (or sensitivity), F1 score, and ROC-AUC score were calculated for the three classifications. Accuracy is the most commonly used performance metric of a classification model that determines the rate of correct classifications. Simply put, accuracy represents a ratio of the correct classifications calculated as the number of correct predictions divided by the number of predictions made. Precision refers to the correctness of a classifier. This number is determined by dividing the number of true positives by the total number of positives (i.e., true positives plus false positives). Recall refers to the sensitivity of a classifier. It is determined by dividing the sum of true positives and false negatives by the total number of true positives. F1 score is the weighted mean of precision and recall. Basically, this score is calculated as two times the product of precision and recall divided by the sum of their parts. ROC-AUC (Receiver Operator Characteristic – Area Under the Curve) is a measure of performance that plots the relationship between the true-positive rate and the false- positive rate for various threshold settings. Its value ranges from 0 to 1. Figure 7 displays plots of ROC curves for the three classifications. As seen from Table 5 the binary classification AD vs. CN gave the highest precision, recall, F1 score, ROC-AUC score, and accuracy, followed by MCIc vs. CN, and then MCIc vs. MCInc. This hierarchy could be justified as a result of the difference in the size of the data available for these four categories as seen in Table 1. Figure 8 represents a graphical illustration for visualizing the improvement in results obtained from the proposed approach.

*Table 4. Accuracy for the three binary classifications*

| Research | Classifier | AD vs. CN | MCIc vs. CN | MCIc vs. MCInc |
|---|---|---|---|---|
| Salvatore et al. (2015) | SVM | 76% | 72% | 66% |
| Pan et al. (2020) | 3D-SENet | 80% | 75% | 57% |
| Pan et al. (2020) | CNN-EL | 84% | 79% | 62% |
| Proposed approach | BoTNet | **91.67%** | **85.22%** | **82.67%** |

*Table 5. Results of different evaluation metrics*

| Classification | Precision | Recall | F1 score | ROC-AUC score | Holdout set accuracy | Validation accuracy |
|---|---|---|---|---|---|---|
| AD vs. CN | 0.907 | 0.895 | 0.897 | 0.89 | 89.5% | 91.67% |
| MCIc vs. CN | 0.88 | 0.87 | 0.87 | 0.86 | 87.28% | 85.22% |
| MCIc vs. MCInc | 0.80 | 0.76 | 0.78 | 0.77 | 76% | 82.67% |

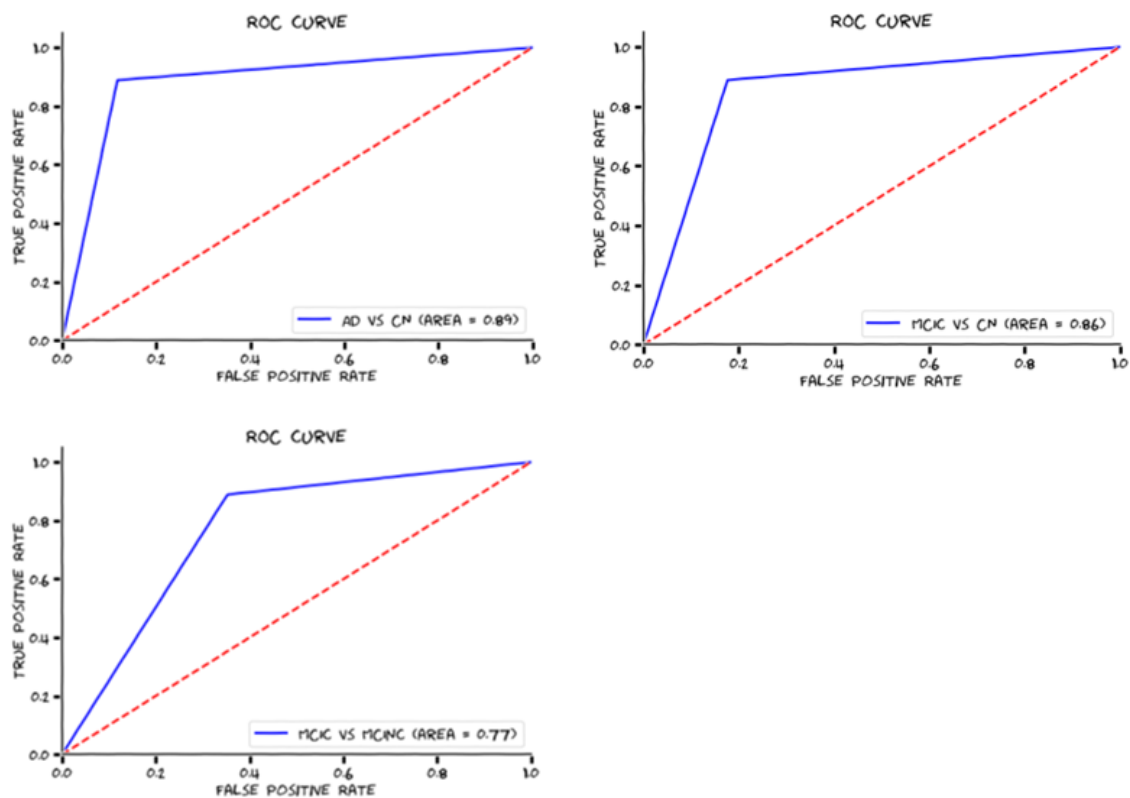

*Figure 7. ROC curves for the three classifications*

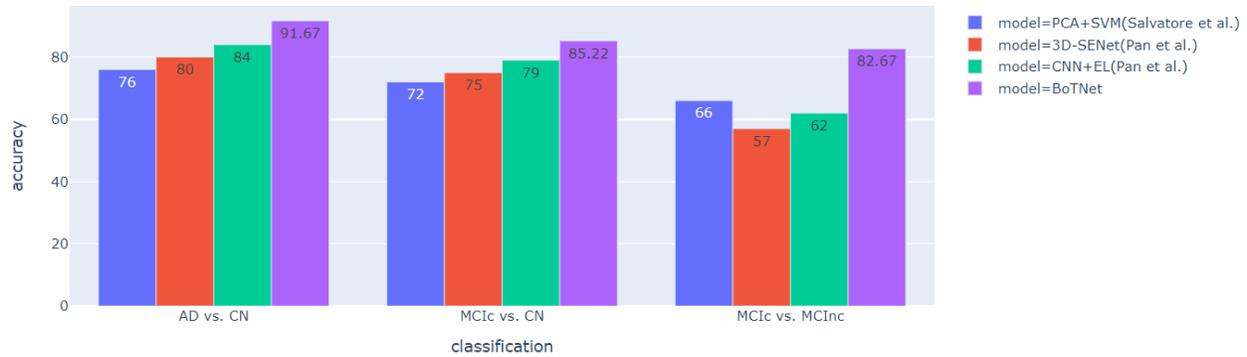

*Figure 8. A bar graph illustrating the results from various research works compared with the one proposed*

## 5. Conclusion

This paper experimented with the use of Bottleneck Transformers (BoTNet) for the early detection of AD by performing three binary classifications between (i) AD and CN, (ii) MCIc and CN, and (iii) MCIc and MCInc. The final accuracies yielded by them were 91.67%, 85.22%, and 82.67% respectively. The advocated methodology showed a significant improvement in the accuracies from previous attempts at these classifications for AD detection. Especially, a remarkable performance enhancement for the classification between MCI converters and non-converters can provide great insight in further understanding what factors contribute to the transition from MCI to AD. Moreover, BoTNet is computationally faster than pure self-attention models like Vision Transformer and just slightly slower than convolution-based models with a significant increase in overall performance. Thus, indicating the importance of hybrid convolution + attention models for computer vision and in particular, for the medical imaging domain.

As a future direction, the proposed approach can be extended to the task of image segmentation for identifying biomarkers that help distinguish between MCI converters and non-converters to facilitate an appropriate treatment plan for the two classes. Furthermore, the prescribed method may be deemed helpful in the early detection of more such diseases as well as be useful in finding their relevant biomarkers.